\documentclass[12pt]{article}
\usepackage{amssymb}
\usepackage{epsfig}
\usepackage{float}
\newcommand{\be}{\begin{equation}}
\newcommand{\ee}{\end{equation}}
\newcommand{\bea}{\begin{eqnarray}}
\newcommand{\eea}{\end{eqnarray}}

\begin{document}
\begin{center}
{\bf A REMARK ON NEUTRINO OSCILLATIONS OBSERVED IN  KamLAND EXPERIMENT}
\end{center}

\begin{center}
S. M. Bilenky
\end{center}
\vspace{0.1cm}
\begin{center}
{\em  Joint Institute
for Nuclear Research, Dubna, R-141980, Russia\\}
\end{center}
\begin{center}
{\em SISSA via Beirut 2-4, Trieste, 34014, Italy\\}
\end{center}

\begin{abstract}
It is demonstrated that the observation of neutrino oscillations in the atmospheric
(K2K) neutrino
experiments and unitarity of the mixing matrix implies that disappearance of the reactor
$\bar\nu_{e}$'s discovered in the KamLAND experiment
is due to $\bar\nu_{e}\to \bar\nu_{\mu}$ {\em and} $\bar\nu_{e}\to \bar\nu_{\tau}$ transitions.
At $\theta_{23}=\pi/4$ the probabilities of these transitions are equal.
\end{abstract}

At the Neutrino 2004 conference a new important result of the KamLAND collaboration was reported:
the significant distortion of the spectrum of the reactor $\bar\nu_{e}$ was observed \cite{kamland}.

As it is well known, in the KamLAND experiment the $\bar\nu_{e}$'s from many reactors
in Japan and Korea are detected via the observation of $e^+$ and $n$
produced in the reaction
\be
\bar\nu_{e}+p \to e^{+}+n
\label{1}
\ee
In the paper \cite{kamland}
it is written " We present an improved measurement of the oscillations
between first two neutrino families....."

In the framework of the standard three neutrino mixing
\be
\nu_{lL}=\sum_{i=1}^{3}U_{li}\nu_{il}
\label{2}
\ee
($U^{\dagger}\,U=1$, $\nu_{i}$ is the field of neutrino with mass $m_{i}$)
we will consider here neutrino oscillations in the KamLAND experiment.

The transition probabilities of neutrino and antineutrinos can be presented in the form
(see \cite{BGG})
\be
{\mathrm P}(\nu_l \to \nu_{l'}) =
|\delta_{ll'} +\sum_{i= 2,3} U_{l' i}  U_{l i}^*
\,~ (e^{- i \Delta m^2_{1i} \frac {L} {2E}} -1)|^2
\label{3}
\ee
and
\be
{\mathrm P}(\bar\nu_l \to \bar\nu_{l'}) =
|\delta_{ll'} +\sum_{i= 2,3} U^*_{l' i}  U_{l i}
\,~ (e^{- i \Delta m^2_{1i} \frac {L} {2E}} -1)|^2
\label{4}
\ee
Here $L$ is the source-detector distance, $E$ is the neutrino energy and
$\Delta m^{2}_{1i}= m^{2}_{i} -m^{2}_{1}$.

From the analysis of the existing neutrino oscillation data two important features of the
neutrino mixing emerged:
\begin{enumerate}
\item
$\Delta m^2_{12}\ll |\Delta m^2_{13}|$
\item
$|U_{e3}|^{2}=\sin^{2}2\theta_{13}\ll 1$
\end{enumerate}

It follows from 1. and 2. that the dominant transition
in the atmospheric range of $L/E$ is $\nu_{\mu}\to \nu_{\tau}$
($\bar\nu_{\mu}\to \bar\nu_{\tau}$). For the $\nu_{\mu}$ ($\bar\nu_{\mu}$)
survival probability from (\ref{3}) and (\ref{4}) we find
\be
{\mathrm P}(\nu_{\mu} \to \nu_{\mu}) ={\mathrm P}(\bar\nu_{\mu} \to \bar\nu_{\mu})
\simeq 1-2 |U_{\mu 3}|^{2}(1- |U_{\mu 3}|^{2})
\,~ (1-cos\Delta m^2_{13} \frac {L} {2E}).
\label{5}
\ee
In the approximation $\sin^{2}2\theta_{13}\to 0$ we have
\be
U_{\mu 3}= \sin\theta_{23},\,~~
U_{\tau 3}= \cos\theta_{23}.
\label{6}
\ee
Thus, in the atmospheric range of $L/E$ we obtain
\be
{\mathrm P}(\nu_{\mu} \to \nu_{\mu}) ={\mathrm P}(\bar\nu_{\mu} \to \bar\nu_{\mu})
\simeq 1-\frac{1}{2}\,~\sin^{2}2\,\theta_{23}
\,~ (1-cos\Delta m^2_{13} \frac {L} {2E}).
\label{7}
\ee
From the analysis of the Super-Kamiokande atmospheric neutrino data
the following best-fit values of
the parameters were found \cite{SK}
\be
\sin^{2}2\,\theta_{23}=1;\,~~\Delta m^2_{13} = 2\cdot 10^{-3}\,\rm{eV}^{2}
\label{8}
\ee
For the probability of the transition $\bar\nu_{e}\to\bar\nu_{l} $
in the KamLAND range of $L/E$ from (\ref{4})
(in the approximation $\sin^{2}2\theta_{13}\to 0$) we find
\be
{\mathrm P}(\bar\nu_e \to \bar\nu_{l}) =
|\delta_{el} + U^*_{l 2}  U_{e 2}
\,~ (e^{- i \Delta m^2_{12} \frac {L} {2E}} -1)|^2
\label{9}
\ee
From this expression we obtain
\be
{\mathrm P}(\bar\nu_{e} \to \bar\nu_{e})
\simeq 1-2 |U_{e2}|^{2}(1- |U_{e2}|^{2})
\,~ (1-cos\Delta m^2_{12} \frac {L} {2E}).
\label{10}
\ee
Further from the unitarity of the mixing matrix we have
\be
U_{e1}\simeq \cos\theta_{12},\,~~
U_{e2}\simeq \sin\theta_{12}.
\label{11}
\ee
Thus, in the
the KamLAND range of $L/E$ the $\bar\nu_{e}$ survival probability is given by the expression
\be
{\mathrm P}(\bar\nu_{e} \to \bar\nu_{e})
\simeq 1-\frac{1}{2}\,~\sin^{2}2\theta_{12}
\,~ (1-cos\Delta m^2_{12} \frac {L} {2E}).
\label{12}
\ee
From the analysis of the latest KamLAND data and solar neutrino data in \cite{kamland}
 it
was found
\be
\Delta m^2_{12} = (8.2 ^{+0.6}_{-0.5})\cdot 10^{-5}\,\rm{eV}^{2};\,
\tan^{2}2\theta_{12}= 0.40 ^{+0.09}_{-0.07}.
\label{13}
\ee
The unitarity of the neutrino mixing matrix implies
\be
{\mathrm P}(\bar\nu_{e} \to \bar\nu_{e}) = 1-({\mathrm P}(\bar\nu_{e} \to \bar\nu_{\mu})+
{\mathrm P}(\bar\nu_{e} \to \bar\nu_{\tau}))
\label{14}
\ee
From (\ref{9}) for the probabilities of the transitions
$\bar\nu_{e} \to \bar\nu_{\mu}$
and $\bar\nu_{e} \to \bar\nu_{\tau}$ we find the following expressions
\be
{\mathrm P}(\bar\nu_{e} \to \bar\nu_{\mu})
\simeq 2 |U_{e2}|^{2}|U_{\mu 2}|^{2}
\,~ (1-\cos\Delta m^2_{12} \frac {L} {2E}).
\label{15}
\ee
and
\be
{\mathrm P}(\bar\nu_{e} \to \bar\nu_{\tau})
\simeq 2 |U_{e2}|^{2}|U_{\tau 2}|^{2}
\,~ (1-\cos\Delta m^2_{12} \frac {L} {2E}).
\label{16}
\ee
The elements $U_{\mu 2}$ and $U_{\tau 2}$ are determined by the angles $\theta_{12}$ and
$\theta_{23}$. In fact, from the unitarity of the mixing matrix we have
\be
\sum_{i=1,2}|U_{\mu i}|^{2}|= \cos^{2}\theta_{23};\,~
\sum_{i=1,2}|U_{\tau i}|^{2}|= \sin^{2}\theta_{23}
\label{17}
\ee
Taking into account that the raws of the mixing matrix must be orthogonal we easily find
\be
U_{\mu 2}= \cos\theta_{23}\,\cos\theta_{12};\,~~
U_{\tau 2}= -\sin\theta_{23}\,\cos\theta_{12}.
\label{18}
\ee
Thus, we have
\be
{\mathrm P}(\bar\nu_{e} \to \bar\nu_{\mu})=
\cos^{2}\theta_{23}\,\frac{1}{2}\,\sin^{2}\theta_{12}
\,~ (1-cos\Delta m^2_{12} \frac {L} {2E}).
\label{19}
\ee
and
\be
{\mathrm P}(\bar\nu_{e} \to \bar\nu_{\tau})=
\sin^{2}\theta_{23}\,\frac{1}{2}\,\sin^{2}\theta_{12}
\,~ (1-cos\Delta m^2_{12} \frac {L} {2E}).
\label{20}
\ee
From (\ref{14}), (\ref{19}) and (\ref{20}) we find the following relations between
transition probabilities in the KamLAND range of $L/E$:
\be
{\mathrm P}(\bar\nu_{e} \to \bar\nu_{\mu})=\cos^{2}\theta_{23}\,
(1-{\mathrm P}(\bar\nu_{e} \to \bar\nu_{e}))
\label{21}
\ee
and
\be
{\mathrm P}(\bar\nu_{e} \to \bar\nu_{\tau})=\sin^{2}\theta_{23}\,
(1-{\mathrm P}(\bar\nu_{e} \to \bar\nu_{e}))
\label{22}
\ee
From these relations it follows that
\be
{\mathrm P}(\bar\nu_{e} \to \bar\nu_{\tau})=\tan^{2}\theta_{23}\,
{\mathrm P}(\bar\nu_{e} \to \bar\nu_{\mu}).
\label{23}
\ee
Therefore the observation of neutrino oscillations in the atmospheric (K2K) experiments and the
unitarity of the neutrino mixing matrix imply that the disappearance of reactor $\nu_{e}$,
observed in the KamLAND experiment, is due to
$\bar\nu_{e} \to \bar\nu_{\mu}$ and $\bar\nu_{e} \to \bar\nu_{\tau}$
transitions. For $\theta_{23}=\pi/2$ (the SK best-fit value) the probabilities
${\mathrm P}(\bar\nu_{e} \to \bar\nu_{\mu})$ and
${\mathrm P}(\bar\nu_{e} \to \bar\nu_{\tau})$ are equal.
\footnote{For arguments in favor of this equality see \cite{Akhmedov}}

Thus, in the leading approximation neutrino oscillations observed in the atmospheric (K2K) experiments
are driven by $\Delta m^2_{13}$ and
are oscillations between second and third neutrino families. Neutrino oscillations observed
in the KamLAND (solar) neutrino
experiments are driven by $\Delta m^2_{12}$. Due to the unitarity of the mixing matrix
all three neutrino families are involved in
the oscillations.

I acknowledge the support of  the Italien Program ``Rientro dei cervelli''.

\end{document}